\documentclass[runningheads]{llncs}
\usepackage{graphicx}
\usepackage{multirow}
\usepackage{float}
\usepackage{tabularx}
\usepackage{booktabs}
\usepackage{pdflscape}
\usepackage{subfig}
\usepackage{todonotes}
\usepackage{tcolorbox}
\usepackage{fixltx2e}
\usepackage{hyperref}

\graphicspath{ {figures/} }

\selectfont

\begin{document}

\title{Evaluating Elements of  \\  Web-based Data Enrichment  for \\ Pseudo-Relevance Feedback Retrieval}

\titlerunning{Evaluating Elements of Web-based Data Enrichment for PRF Retrieval}

\pagestyle{plain}

\author{Timo Breuer \and Melanie Pest \and Philipp Schaer}
\authorrunning{Breuer et al.}
\institute{TH K\"oln (University of Applied Sciences)
\newline
\email{firstname.lastname@th-koeln.de}}

\maketitle     
\begin{abstract}
In this work, we analyze a pseudo-relevance retrieval method based on the results of web search engines. By enriching topics with text data from web search engine result pages and linked contents, we train topic-specific and cost-efficient classifiers that can be used to search test collections for relevant documents. Building upon attempts initially made at TREC Common Core 2018 by Grossman and Cormack, we address questions of system performance over time considering different search engines, queries, and test collections. Our experimental results show how and to which extent the considered components affect the retrieval performance. Overall, the analyzed method is robust in terms of average retrieval performance and a promising way to use web content for the data enrichment of relevance feedback methods.

\end{abstract}

\section{Introduction}

Modern web search engines provide access to rich text data sources in search engine result pages (SERPs) and linked web page contents. In the effort of being comprehensive, search results should be diverse but also up-to-date. Although the search results themselves are intended for presentation to search engine users, they could be used to supplement automatic searching in a text collection, e.g., in the form of data enriched relevance feedback methods.

This work builds upon attempts that were first made by Grossman and Cormack (GC) at TREC Common Core18. The authors exploit the results of web search engines to enrich a pseudo-relevance retrieval method that ranks documents of the TREC Washington Post Corpus. More specifically, multiple pseudo-relevance classifiers are individually trained for each topic of the test collection. Each of these classifiers is based on training data retrieved from texts of scraped SERPs, which, in turn, depend on the query of the related topic. Depending on the topic, results of specific requests to web search engines may be subject to strong time dependencies, e.g., when related to breaking news. Likewise, the geolocation or algorithmic changes in retrieval techniques and snippet generation can influence search results or the way they are presented. Thus, it is of special interest to investigate the reliability of this retrieval method that relies on ephemeral and constantly changing web content. On the other hand, the analyzed approach is a promising way to build cost-efficient classifiers that find relevant documents. % in any collection.

Our contributions are twofold. First, we address the following research questions: \textbf{RQ1} \textit{How do the components of the workflow, i.e., the query formulation and the web search engine, affect the system performance over time?} and \textbf{RQ2} \textit{To which extent are the original effects present in different contexts, i.e., with other newswire test collections?} Second, we provide an open-source implementation of the analyzed approach and make the scraped web contents and system runs available for follow-up studies. The remainder is structured as follows. Section 2 contains the related work. Section 3 explains how data enriched topics are used to train pseudo-relevance classifiers and provides details about the analyzed workflow components and the corresponding modifications to the experiments. Section 4 presents the experimental results that are based on reproducibility measures of system-oriented IR experiments. Section 5 concludes.

\section{Related Work}

Pointwise learning to rank (LTR) algorithms directly apply supervised machine learning approaches for predicting the relevance of single documents~\cite{DBLP:journals/ftir/Liu09}. Usually, these approaches require training data with explicit class labels that are costly due to editorial efforts. Relevance feedback algorithms form another body of research in information retrieval (IR) literature correlated to pointwise LTR approaches~\cite{DBLP:journals/ker/RuthvenL03}. Here, relevance feedback based on previously retrieved documents is exploited to improve the final ranking results. Pseudo-relevance feedback (PRF) algorithms - a specific type of relevance feedback - omit editorial labeling by assuming the top-$k$ previously retrieved documents to be relevant~\cite{DBLP:journals/jd/CroftH79}. Deriving training data with PRF mechanisms and applying it to LTR algorithms requires explicit positive and negative training samples. Raman et al. successfully showed that the assignment of explicit pseudo-irrelevant documents could improve the retrieval effectiveness~\cite{DBLP:conf/ecir/RamanUBB10}. Following a string of different text classification approaches based on machine learning, including spam-filtering~\cite{DBLP:journals/ir/CormackSC11} and cross-collection relevance classification~\cite{DBLP:conf/trec/GrossmanC17}, GC propose to train a logistic regression classifier with pseudo-ir/relevant tfidf-features from SERPs and linked web pages~\cite{DBLP:conf/trec/GrossmanC18}. Similarly, Nallapati investigates different combinations of tf, idf, and combined statistics with discriminative classifiers~\cite{DBLP:conf/sigir/Nallapati04}. Xu and Akella investigate the relevance classification based on the combination of relevance feedback with a logistic regression classifier~\cite{DBLP:conf/sigir/XuA08}.

Relying on the web as a large external corpus for query expansions has been extensively exploited by the top-performing 
groups at TREC Robust in 2004 and 2005~\cite{DBLP:conf/trec/Voorhees04b,DBLP:conf/trec/Voorhees05a} or as part of TREC Microblog~\cite{DBLP:journals/ijws/BandyopadhyayGMM12}. Kwok et al. showed that web-assisted query expansions could improve the retrieval effectiveness, especially for short queries or queries having only a semantic relation~\cite{DBLP:conf/airs/KwokGD05}. Similarly, Diaz and Metzler showed that query expansions from external corpora improve the mean average precision, especially when the external corpus is larger than the target collection~\cite{DBLP:conf/sigir/DiazM06}. As part of their experimental setups, they include web documents from Yahoo! web corpus. Like web documents, Wikipedia offers a resource for PRF via query expansions, as investigated by Xu et al.~\cite{DBLP:conf/sigir/XuJW09} and Li et al.~\cite{DBLP:conf/sigir/LiLHC07}. Yu et al. showed that PRF techniques are more successful when considering the semantic structure of web pages~\cite{DBLP:conf/www/YuCWM03}. When relying on hypertext documents, it is important to consider the markup removal before indexing, as shown by Roy et al.~\cite{DBLP:journals/jdiq/RoyMG18}. Otherwise, markup text affects the final results and the reproducibility.

The previously mentioned approach by GC is based on a PRF mechanism, and the principle of routing runs~\cite{DBLP:conf/trec/GrossmanC18}. As defined by Robertson and Callan~\cite{robertson_callan}, a routing profile is constructed from existing judgments and used for assessing the relevance of documents. GC propose to derive this profile automatically by generating training data from SERPs scraped for specific topics. Subsequently, the profile ranks documents of the test collection. More specifically, the training data is retrieved from Google SERPs, and the resulting profile ranks the TREC Washington Post Corpus (Core18). GC submitted two alternative runs. In both cases, queries are concatenations of topic titles and descriptions. In order to derive the first run variant \texttt{uwmrg}, the entire content of web pages corresponding to the URLs of the SERP is scraped.  In their second submission  \texttt{uwmrgx}, GC propose using only the snippets from SERPs instead of scraping complete web pages. After having retrieved results for all topics of a test collection, text data is prepared as training samples and used for modeling a routing profile that ranks documents from Core18 scored by their probability of being relevant.

\section{Approach}

The workflow is concise and straightforward. Compressed files are extracted with GNU tools \texttt{tar} and \texttt{gzip}. Likewise, the required web content is scraped and freed from markup artifacts with the help of the \texttt{BeautifulSoup} Python package. The preprocessing removes punctuation and stop words, includes stemming (PorterStemmer), and is implemented with the help of the \texttt{nltk} Python package. The preprocessed text data of the scraped web content is used to derive a term-document matrix with tfidf-weights for which we use a sublinear scaling of term frequencies. For this purpose, we make use of the \texttt{TfidfVectorizer} of the \texttt{scikit-learn} Python package. This term-document matrix is also used for generating tfidf-feature vectors of documents from the test collection. 

% \begin{figure}[t!]
% \centering
% \includegraphics[width=0.7\textwidth]{uwmrg-repro.pdf}
% \caption{Visualization of the workflow proposed by Grossman and Cormack~\cite{DBLP:conf/trec/GrossmanC18}.}
% \label{workflow}
% \end{figure}

Training data is stored in \texttt{SVMlight} format. Class assignments of training features (to positive and negative samples) are based on a one-vs-rest principle. Depending on the topic, positive samples will be retrieved with the corresponding title (and description), while scraped results of other topics serve as negative samples. After training data and test collection features have been prepared, a model is trained for each topic. In this context, a logistic regression classifier is used. Again, we rely on \texttt{scikit-learn} when implementing the \texttt{LogisticRegression} classifier. Adhering to the details given by GC we set the tolerance parameter to $0.0001$ (default in \texttt{scikit-learn}) and the number of maximal iterations to $200,000$. Subsequently, documents are ranked by their likelihood of being relevant. The 10,000 first documents for each topic contribute to the final system run. For more specific and technical details, we refer the reader to our GitHub repository\footnote{\href{https://github.com/irgroup/clef2021-web-prf/}{\url{https://github.com/irgroup/clef2021-web-prf/}}}.

\subsection{Modifications in the Experiments}
\label{sec:determinants}

As pointed out earlier, we are interested in the robustness of the introduced approach. \textbf{RQ1} includes multiple aspects that are addressed as follows. In our case, the \textit{test of time} addresses possibly different web contents in comparison to the original experiments. SERPs are subject to a strong time dependency and the returned URLs, including how they are presented, change over time. In order to investigate the influence of time, we compare results with various time differences. First, we compare our results scraped in 2020 to the original experiment from 2018. On a more granular level, we compare results based on Core18 that were scraped every second day for 12 days in June 2020.

The original results were based on training data scraped from Google SERPs only. We investigate the influence of the \textit{web search engine} by contrasting it with an alternative. In our study, we scrape training data from Google and DuckDuckGo, since the latter states not to include any personalized information except for optional localization contexts\footnote{\href{https://spreadprivacy.com/why-use-duckduckgo-instead-of-google/}{\url{https://spreadprivacy.com/why-use-duckduckgo-instead-of-google/}}, \\ accessed: May 3rd, 2021}.

In the original experiment, the \textit{queries} were made of concatenated topic titles and descriptions. However, we argue it might be interesting to contrast this approach with queries made of the topic title only. Users tend to formulate short queries for many web search tasks instead of thoroughly and explicitly formulating their search intents. Thus, we include both short (\texttt{title}) and longer (\texttt{title+desc}) queries in our experimental setup.

Addressing \textbf{RQ2}, we extend the investigations by considering four different \textit{test collections} in total.  Originally, runs were derived from the TREC Washington Post Corpus. We investigate and compare results retrieved from TREC Washington Post Corpus (Core18), New York Times Annotated Corpus (Core17), the AQUAINT collection (Robust05), and TREC disks 4 and 5 (minus the \textit{Congressional Record}) (Robust04). All test collections contain newswire documents.

Finally, it has to be noted that SERPs are affected by personalization, as shown by Hannak et al.~\cite{DBLP:conf/www/HannakSKKLMW13}. One major influence of personalization is the \textit{geolocation}~\cite{DBLP:conf/www/YiRL09}. Throughout our experiments, we do not vary the geolocation parameter when querying the web search engines. We assume the original results to be retrieved with a Canadian geolocation parameter\footnote{GC are affiliated with the University of Waterloo in Canada}, and we also use an English language code when querying the web search engine. Likewise, we keep the classifier's parameterization and how training data is preprocessed fixed to minimize their influence. As part of this study, we limit the training data to texts extracted from either web pages (\texttt{uwmrg}) or snippets (\texttt{uwmrgx}) corresponding to the ten first search results for each topic. While modern SERPs offer more comprehensive results than ``ten blue links'', we do not include other SERP sections like related queries/searches, entity cards, or advertisements. In the future, it might be interesting to include the influence of these elements.

\section{Experimental Results}

Our experimental evaluations are made with the toolkit \texttt{repro\_eval}~\cite{DBLP:conf/ecir/BreuerFMS21} that implements reproducibility measures of system-oriented IR experiments~\cite{DBLP:conf/sigir/Breuer0FMSSS20}. More specifically, we analyze the regression tests made with our reimplementation compared to the original results by GC. % concerning the correlation between the ordering of documents, the reproduced effectiveness and overall effects. 
Even though this is not a reproducibility study in the true sense of the word, we see these measures as the right tool to answer how much the results vary over time considering the analyzed modifications. For some measures, a baseline and an advanced run are required. In the following, we consider \texttt{uwmrgx} as the baseline run based on SERP snippets and \texttt{uwmrg} as the websites' full text based advanced version of it (since the results were more effective in the original experiment)\footnote{Most Tables and Figures contain results instantiated with nDCG. For results instantiated with other measures, please have a look at the online appendix at \href{https://github.com/irgroup/clef2021-web-prf/blob/master/doc/appendix.pdf}{\url{https://github.com/irgroup/clef2021-web-prf/blob/master/doc/appendix.pdf}}}.

\subsubsection{RQ1: Influence of the Web Search Engine and the Query Formulation}

Table~\ref{tab:core18} shows the reimplemented results derived from the original test collection (\texttt{c18}) under variations of the web search engine (\texttt{g}: Google; \texttt{d}: DuckDuckGo) and the query formulation (\texttt{t}: Topic title only; \texttt{td}: Topic title and description). To the best of our knowledge, the run-type \texttt{c18\_g\_td} exactly corresponds to the original configurations. With regard to the nDCG scores, the reproduced results are fairly good. When using DuckDuckGo for retrieving web results (\texttt{c18\_d\_td} and \texttt{c18\_d\_t}), the reimplemented baseline scores are slightly higher than the original results. Even though the reimplemented advanced nDCG scores do not exceed the original scores in each of the four cases, we consider our reimplementations a good starting point for further investigations.

\begin{table}[t!]
\begin{center}
\captionof{table}{Results of reproduced baseline and advanced runs derived from Core18.} 
\label{tab:core18}
\begin{tabularx}{\textwidth}{l|X|X|X|X||X|X|X|X}
\toprule
\multicolumn{1}{c}{} & \multicolumn{4}{c}{\texttt{uwmrgx} (baseline run)} & \multicolumn{4}{c}{\texttt{uwmrg}
(advanced run)} \\
\midrule
Run & nDCG & KTU & RBO & RMSE & nDCG & KTU & RBO & RMSE \\
\midrule
GC~\cite{DBLP:conf/trec/GrossmanC18} & 0.5306 & 1 & 1 & 0 & 0.5822 & 1 & 1 & 0 \\
\midrule
\texttt{c18\_g\_td}  &  0.5325  &  0.0052  &  0.2252  &  0.1420  &  0.5713  &  0.0071  &  0.3590  &  0.0885 \\
\texttt{c18\_g\_t}  &  0.5024  &  0.0024  &  0.2223  &  0.1697  &  0.5666  &  -0.0030  &  0.3316  &  0.0893 \\
\texttt{c18\_d\_td}  &  0.5735  &  -0.0024  &  0.2205  &  0.1678  &  0.5633  &  -0.0001  &  0.3558  &  0.1014 \\
\texttt{c18\_d\_t}  &  0.5458  &  -0.0020  &  0.1897  &  0.1387  &  0.5668  &  -0.0020  &  0.3357  &  0.1083 \\
\bottomrule
\end{tabularx}
\end{center}
\end{table}
\vspace{5pt}

\begin{figure}[t!]
    \centering
    \subfloat{{\includegraphics[width=4.25cm]{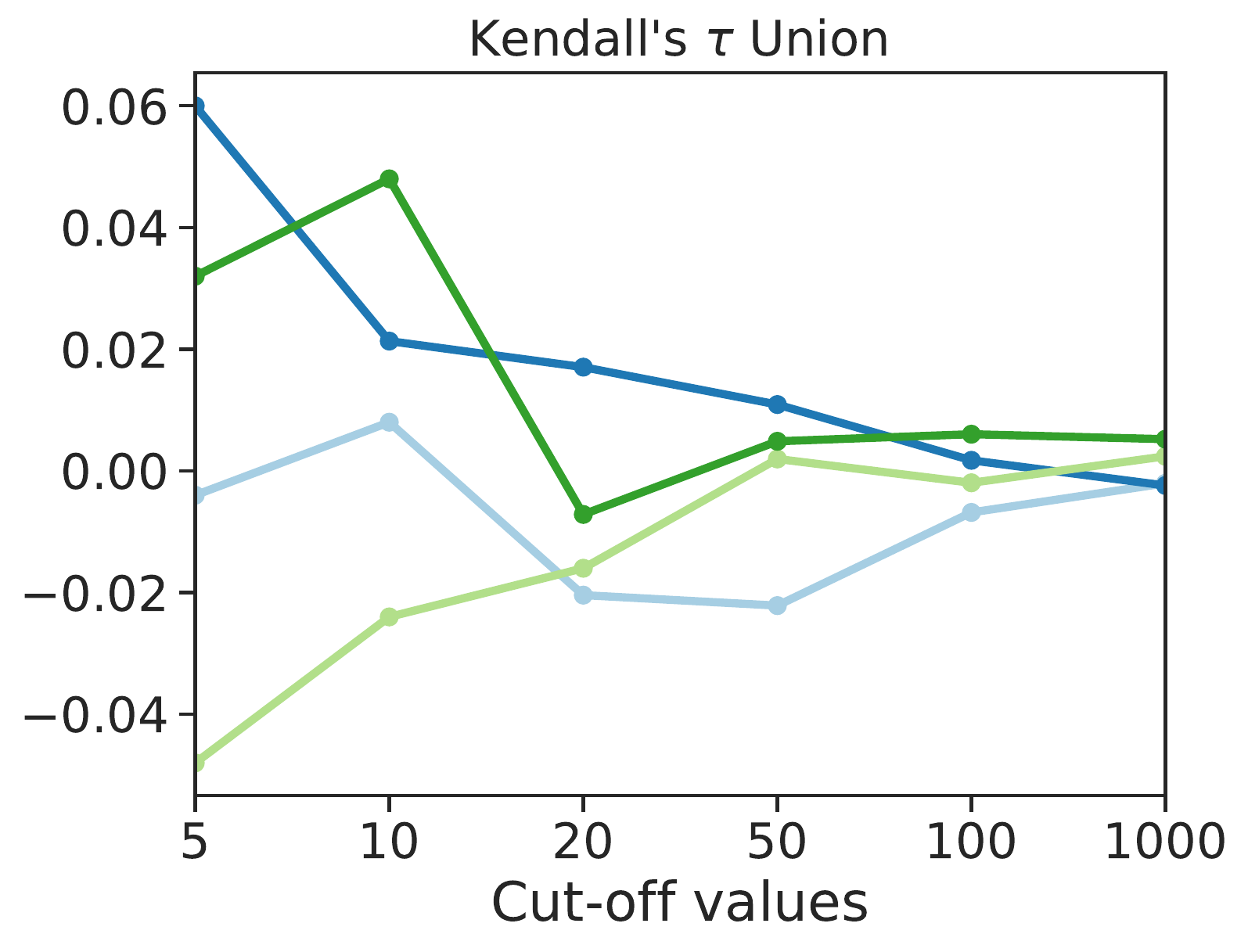}}}
    \subfloat{{\includegraphics[width=4.1cm]{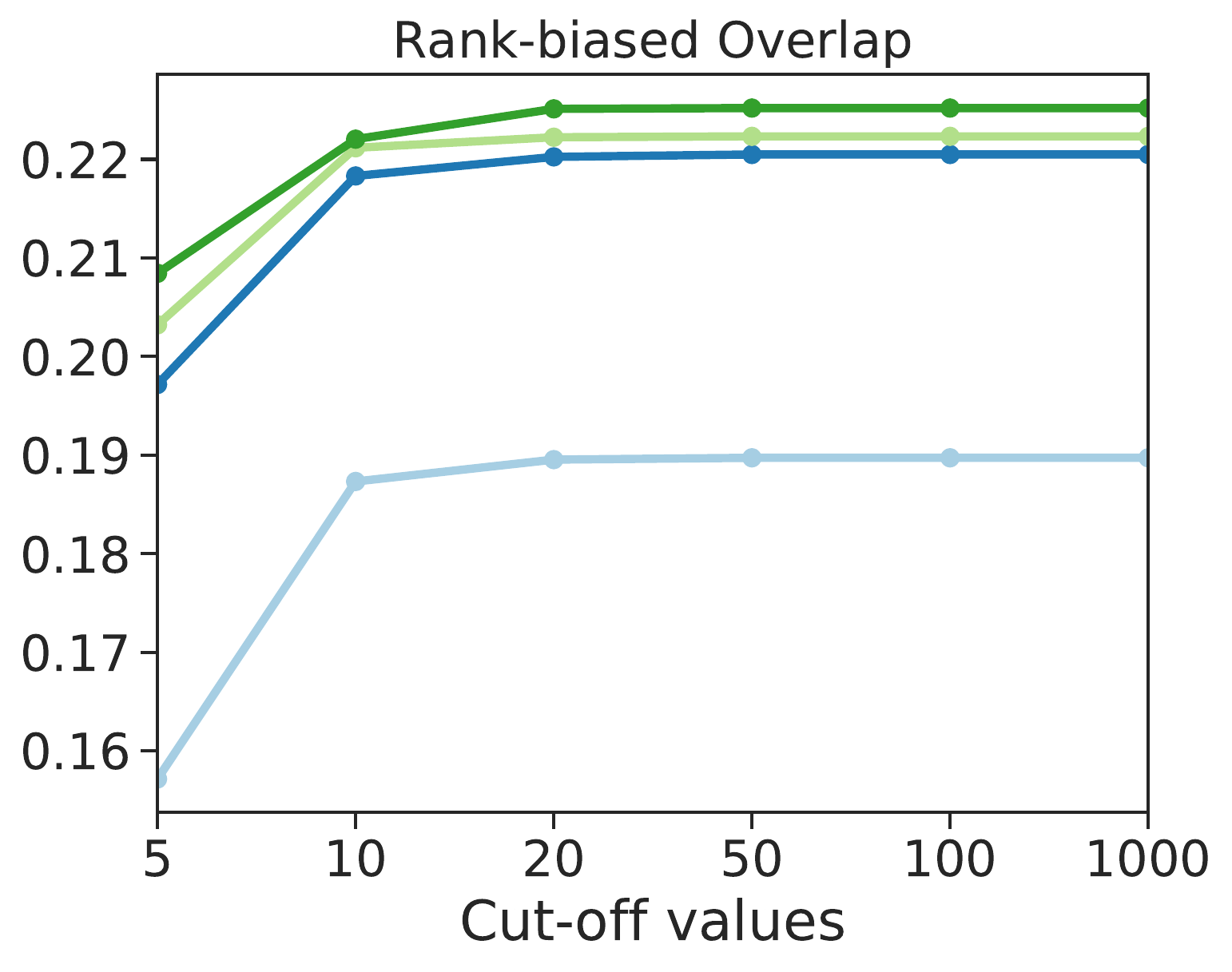}}}
    \subfloat{{\includegraphics[width=4.1cm]{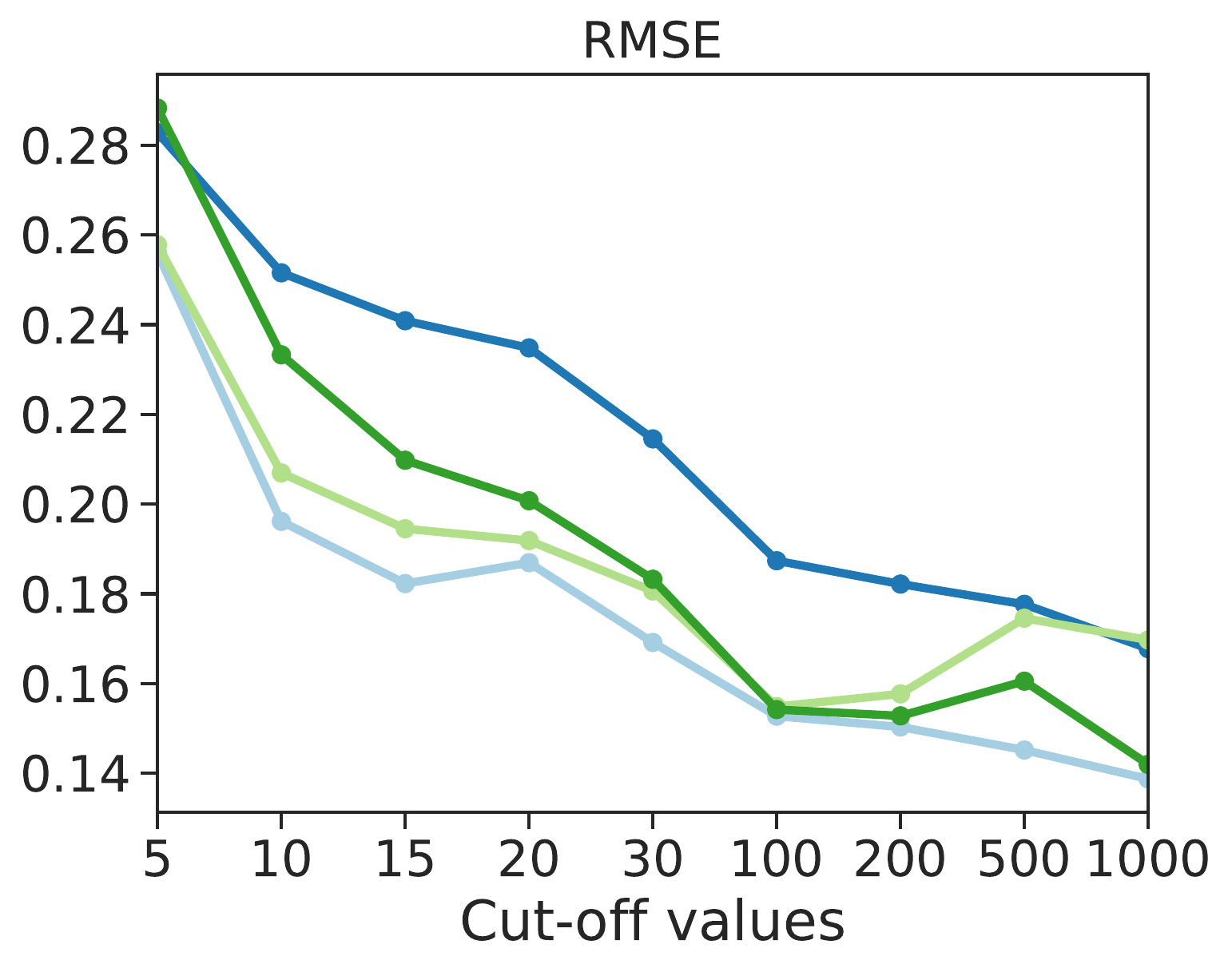}}}
    
    \subfloat{{\includegraphics[width=12cm]{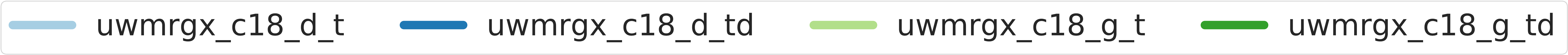}}}
    \caption{Kendall's $\tau$ Union, Rank-biased Overlap, and the Root-Mean-Square-Error of the reproduced baseline \texttt{uwmrgx} averaged across the topics of Core18.}
    \label{fig:ktu_rbo_rmse}
\end{figure}
 
At the most specific level, we compare the reimplementations with the help of Kendall's $\tau$ Union (KTU) and the Rank-Biased Overlap (RBO)~\cite{DBLP:journals/tois/WebberMZ10}. Table~\ref{tab:core18} and Figure~\ref{fig:ktu_rbo_rmse} compare the KTU scores of the reproduced results. In contrast to Kendall's  $\tau$, KTU lowers the restriction of comparing the actual documents by considering lists of ranks instead~\cite{DBLP:conf/sigir/Breuer0FMSSS20}. However, it is still a very strict measure, and as the results show, there is no correlation between the original and reimplemented rankings, neither for the baseline nor for the advanced runs. In addition, Figure~\ref{fig:ktu_rbo_rmse} shows the KTU scores across the different cut-off ranks. Likewise, the ordering of documents is low correlated across the different ranks.

The RBO measure can be used to compare lists with indefinite lengths and possibly different documents. Table~\ref{tab:core18} and Figure~\ref{fig:ktu_rbo_rmse} show comparisons of the reimplemented RBO scores. The rankings based on training data from DuckDuckGo combined with short queries (\texttt{uwmrgx\_c18\_d\_t}) result in lower RBO scores in comparison to the other baseline runs. The other three reimplementations of the baseline do not differ much across the cut-off ranks in terms of RBO. Similarly, there are differences between the reimplemented advanced runs. Runs based on queries with title and description achieve slightly higher RBO scores in comparison to the reimplementations with title-only queries. When comparing the advanced reimplementations to the baselines, there are higher RBO scores for the advanced runs (e.g. RBO\textsubscript{\texttt{uwmrgx\_c18\_g\_td}} $0.2252$ vs. RBO\textsubscript{\texttt{uwmrg\_c18\_g\_td}} $0.3590$). Combining Google and title-only queries results in the lowest RBO scores for the advanced runs, whereas in the case of baseline runs, it does not differ much from those runs based on queries with title and description.

At the level of effectiveness, the Root-Mean-Square-Error (RMSE), which is reported in Table~\ref{tab:core18} and Figure~\ref{fig:ktu_rbo_rmse}, measures the closeness between the topic score distributions~\cite{DBLP:conf/sigir/Breuer0FMSSS20}. As a rule of thumb, the closer the RMSE to a value of $0$, the smaller is the deviation. Interestingly, the baseline run \texttt{uwmrgx\_c18\_d\_t} achieves the lowest RMSE ($0.1387$), despite its low correlation of document orderings in terms of RBO. With regard to the advanced reimplementations, \texttt{uwmrg\_c18\_g\_td} (most similar to the original experiment) achieves the lowest RMSE@$1000$ ($0.0885$). Figure~\ref{fig:ktu_rbo_rmse} illustrates the decreasing RMSE scores with increasing cut-off values. For both baseline and advanced runs, there are almost consistently lower RMSE across the cut-off ranks. Similar to higher RBO scores of the advanced runs, there is a lower RMSE compared to those of the baselines.

Additionally, we conducted t-tests between the runs of our reimplementations to test for significant differences between the search engines or the query formulations (cf. Table \ref{tab:overall_effects}). For each test collection, we either compared runs from different search engines with the same query (significant differences denoted with $\dagger$) or compared runs with different queries but the same search engine (significant differences denoted with $*$). There are slightly lower absolute values but mostly insignificant differences with regard to the different query types. However, when using title-only queries, there are often significant differences between the search engines for those runs based on the snippets' texts only (\texttt{uwmrgx}). 

\subsubsection{RQ1: Influence of Time}

\begin{figure}[t!]
\centering
\includegraphics[width=\textwidth]{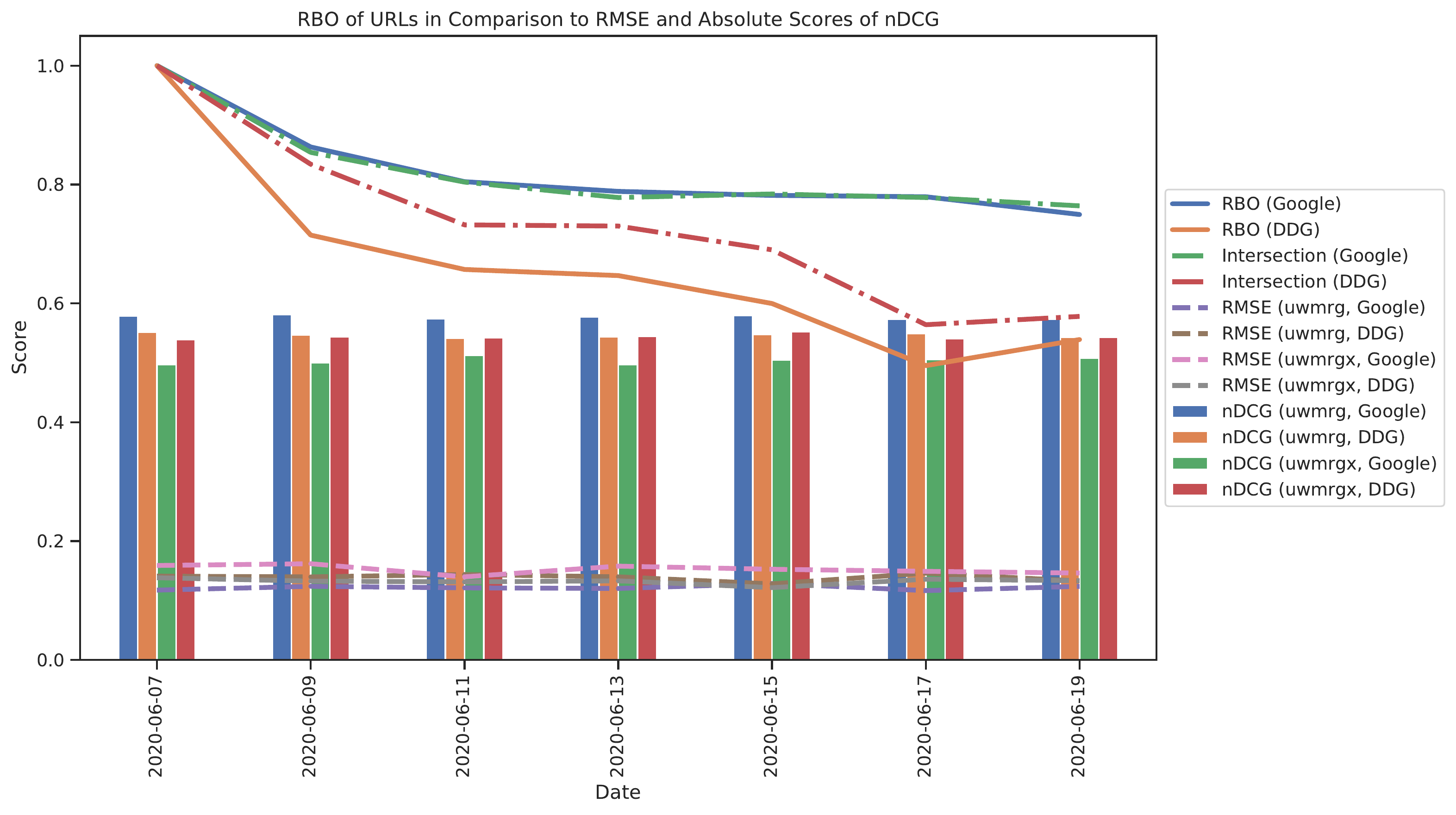}
\caption{RBO scores and relative amount of intersecting URLs in comparison to the nDCG and RMSE scores.}
\label{fig:time_series}
\end{figure}

In Table~\ref{tab:core18}, we compare our reimplementations to the original results of approximately two years ago. However, as pointed out earlier, web content and especially SERPs are subject to several influences, and like the web content itself, they change frequently. Thus, it is worth investigating the robustness of our reimplementations on a more granular level. For this purpose, we retrieved training data from both web search engines for 12 days, starting on June 7th, 2020. Figure~\ref{fig:time_series} shows the RBO, and the intersections between the URLs scraped at every second day compared to those scraped at the beginning on June 7th, 2020. Additionally, Figure~\ref{fig:time_series} includes the absolute nDCG scores and the RMSE scores of the reproduced baseline runs. While the RBO scores decrease over time, the nDCG and RMSE scores are robust with slight variations. We find a strong correlation between the RBO scores and the number of intersecting URLs in the search result lists\footnote{Pearson's $r=0.9747$ and $p=0.0002$} - the lower the RBO, the fewer URLs are in both SERP lists from different days. While it is out of this study's scope to reach any definitive conclusions, we see that the SERP's actual search results (and their URL orders) do not have to be the same as in the original experiment to reproduce the system performance and the effectiveness. Under the consideration of this ``bag of words'' approach, we assume that the results can be reproduced with different web search results having a similar vocabulary or tfidf-features that resemble those used to train the classifiers in the original experiments.

\subsubsection{RQ2: Other Test Collections}

\begin{figure}[t!]
\centering
\includegraphics[width=\textwidth]{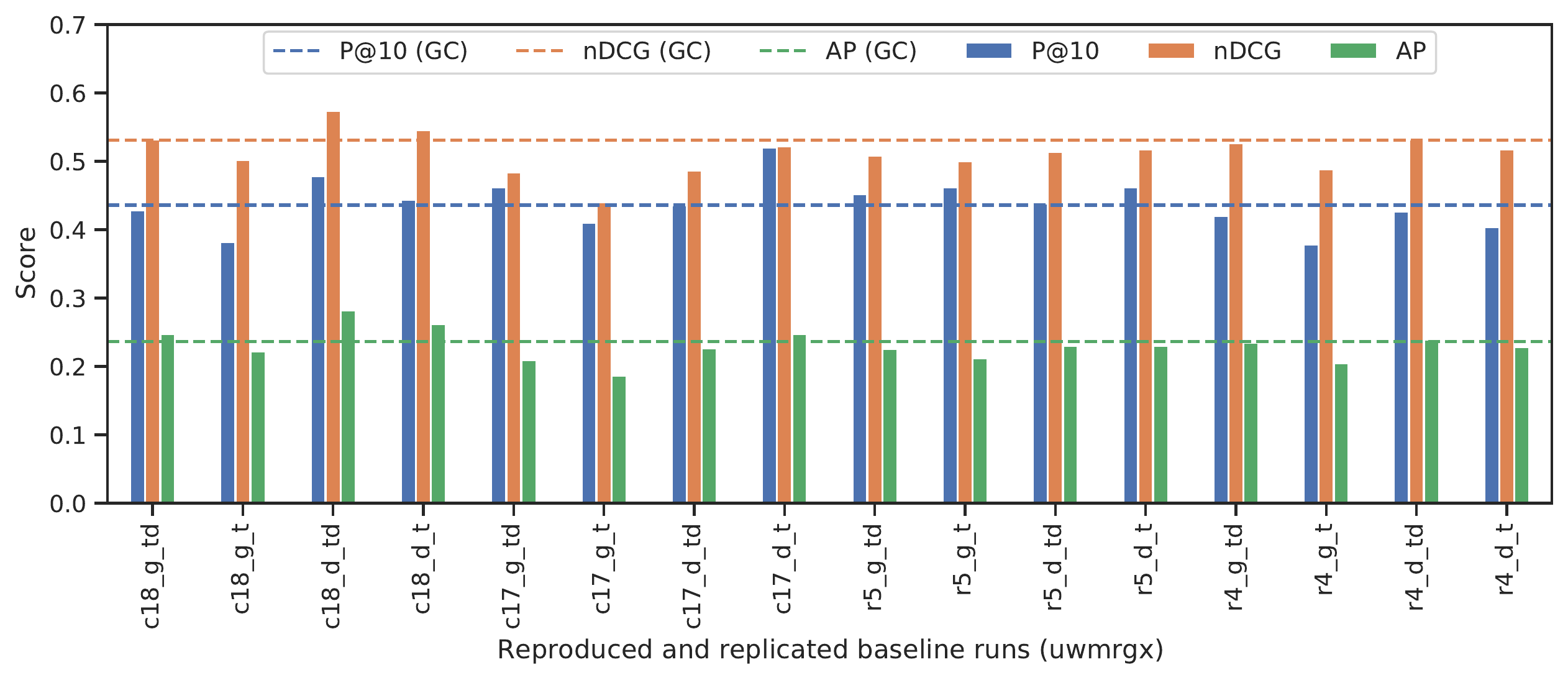}
\caption{Absolute scores of reproduced and replicated baseline runs derived from Core18 (\texttt{c18}), Core17 (\texttt{c17}), Robust04 (\texttt{r4}) and Robust05 (\texttt{r5}).} 
\label{absolute_scores}
\end{figure}

In the following, we evaluate the reimplementations by replacing the target collection. Figure~\ref{absolute_scores} shows the AP, nDCG, and P@10 scores of the baseline runs derived from four different newswire test collections with variations of the query type and web search engine. Our un-/paired t-tests (between the original runs and the reimplementations) show significant differences in very few cases (cf. online appendix), which is an indicator of robustness from a statistical perspective.
%When comparing the results by the two different query types, there are slightly lower absolute scores in most cases. 
% For the replicated results, it is not possible to compare the replications to the original runs in terms of KTU, RBO, and RMSE. Since the replicated runs are derived from a different test collection, they contain different documents for possibly different topics. 
When replacing the target collection, it is impossible to compare KTU, RBO, and RMSE since the runs contain different documents for possibly different topics. In this case, the experiment can be evaluated at the level of overall effects. Here, the Effect Ratio (ER) and the Delta Relative Improvement (DRI) measure the effects between the baseline and advanced runs~\cite{DBLP:conf/sigir/Breuer0FMSSS20}. Perfectly replicated effects are equal to ER $1$, whereas lower and higher scores than $1$ indicate weaker or stronger effects, respectively, than in the original experiment. The DRI complements the ER by considering the absolute scores of the effects. In this case, perfect replication is equal to DRI $0$. Likewise, lower and higher scores indicate weaker or stronger effects, respectively. Table~\ref{tab:overall_effects} shows the overall effects instantiated with nDCG. % and AP of all different run types.

Comparing both search engines, the reproduced and replicated overall effects tend to be higher in ER for training data retrieved with Google. Especially, training data from Google with title-only queries (\texttt{g\_t}) results in ER~$>1$ across all test collections. This can be explained by lower replicability scores for baseline runs, while the advanced runs resemble the original scores fairly well. For instance, the \texttt{uwmrg\_r5\_g\_t} run achieves nDCG $0.5865$ while the corresponding baseline run \texttt{uwmrgx\_r5\_g\_t} results in  nDCG $0.5003$. Consequently, this results in ER $1.6712$ indicating larger effects between the baseline and advanced version than in the original experiment. For results based on training data from DuckDuckGo there are weaker overall effects with ER~$<1$ for each combination of test collection and query type. In most cases, the baseline scores are higher than the corresponding counterparts based on Google results, whereas the advanced scores are lower than those from Google or the original experiments. For instance, \texttt{c18\_d\_td} results in ER\textsubscript{nDCG} $-0.1985$. Here, the baseline scores are higher than those of the advanced versions.

Another way to illustrate the overall effects is to plot the DRI\textsubscript{nDCG} against ER\textsubscript{nDCG} for runs based on training data from Google or DuckDuckGo. In general, it can be said that the closer a point to (ER $1$, DRI $0$), the better the replication. The colors distinct runs with title-only queries (blue) from title and description queries (green). As can be seen, for Google, the data points are distributed over the second and fourth quadrants, whereas for DuckDuckGo all data points are in the second quadrant. With regard to Google, all title-only data points are in the fourth quadrant. This confirms the previous interpretations: training data from Google with title-only queries results in stronger overall effects than in the original experiment.

\begin{table}[ht]
\captionof{table}{Overall effects (with nDCG) of the search engines (SE) and queries (Q) consisting of \texttt{title} (blue) and \texttt{title+desc} (green) of different run versions. $\dagger$ and $*$ denote significant differences ($p<0.05$) between SE and Q, respectively.}
\label{tab:overall_effects}
\begin{minipage}[t]{0.525\linewidth}\centering
\resizebox{\columnwidth}{!}{
\begin{tabularx}{\textwidth}{l|X|X|X|X}
\toprule
\multicolumn{1}{c}{}  &  \multicolumn{2}{c}{nDCG} & \multicolumn{2}{c}{Overall Effects} \\
\midrule
Run & \texttt{uwmrgx} & \texttt{uwmrg} & DRI & ER \\
\midrule
GC~\cite{DBLP:conf/trec/GrossmanC18} & 0.5306 & 0.5822 & 0 & 1 \\
\midrule
\texttt{c18\_g\_td}  &  $0.5325^{\dagger}$  &  0.5713  &  0.0242  &  0.7538  \\
\texttt{c18\_g\_t}  &  $0.5024{^\dagger}$  &  0.5666  &  -0.0305  &  1.2445   \\
\texttt{c18\_d\_td}  &  $0.5735{^\dagger}$  &  0.5633  &  0.1150  &  -0.1985  \\
\texttt{c18\_d\_t}  &  $0.5458{^\dagger}$  &  0.5668  &  0.0587  &  0.4067   \\
\midrule
\texttt{c17\_g\_td}  &  0.4836  &  0.5047  &  0.0534  &  0.4107  \\
\texttt{c17\_g\_t}  &  $0.4404^{\dagger}$  &  0.5313  &  -0.1093  &  1.7637   \\
\texttt{c17\_d\_td}  &  0.4870  &  0.5201  &  0.0291  &  0.6425   \\
\texttt{c17\_d\_t}  &  $0.5223^{\dagger}$  &  0.5279  &  0.0864  &  0.1090   \\
\midrule
\texttt{r5\_g\_td}  &  0.5088  &  0.5613  &  -0.0061  &  1.0192  \\
\texttt{r5\_g\_t}  &  0.5003  &  $0.5865^{\dagger}$  &  -0.0750  &  1.6712   \\
\texttt{r5\_d\_td}  &  0.5134  &  0.5295  &  0.0659  &  0.3110   \\
\texttt{r5\_d\_t}  &  0.5175  &  $0.5509^{\dagger}$  &  0.0325  &  0.6486   \\
\midrule
\texttt{r4\_g\_td}  &  $0.5266^{*}$   &  $0.5357^{*}$   &  0.0798  &  0.1772  \\
\texttt{r4\_g\_t}  &  $0.4886^{\dagger*}$  &  $0.5509^{*}$   &  -0.0304  &  1.2091  \\
\texttt{r4\_d\_td}  &  $0.5317^{*}$   &  0.5376  &  0.0861  &  0.1134  \\
\texttt{r4\_d\_t} &  $0.5171^{\dagger*}$  &  0.5411  &  0.0508  &  0.4651  \\
\bottomrule
\end{tabularx}
}
\end{minipage}
\begin{minipage}[t]{0.45\linewidth}
    \centering
    \includegraphics[width=6cm]{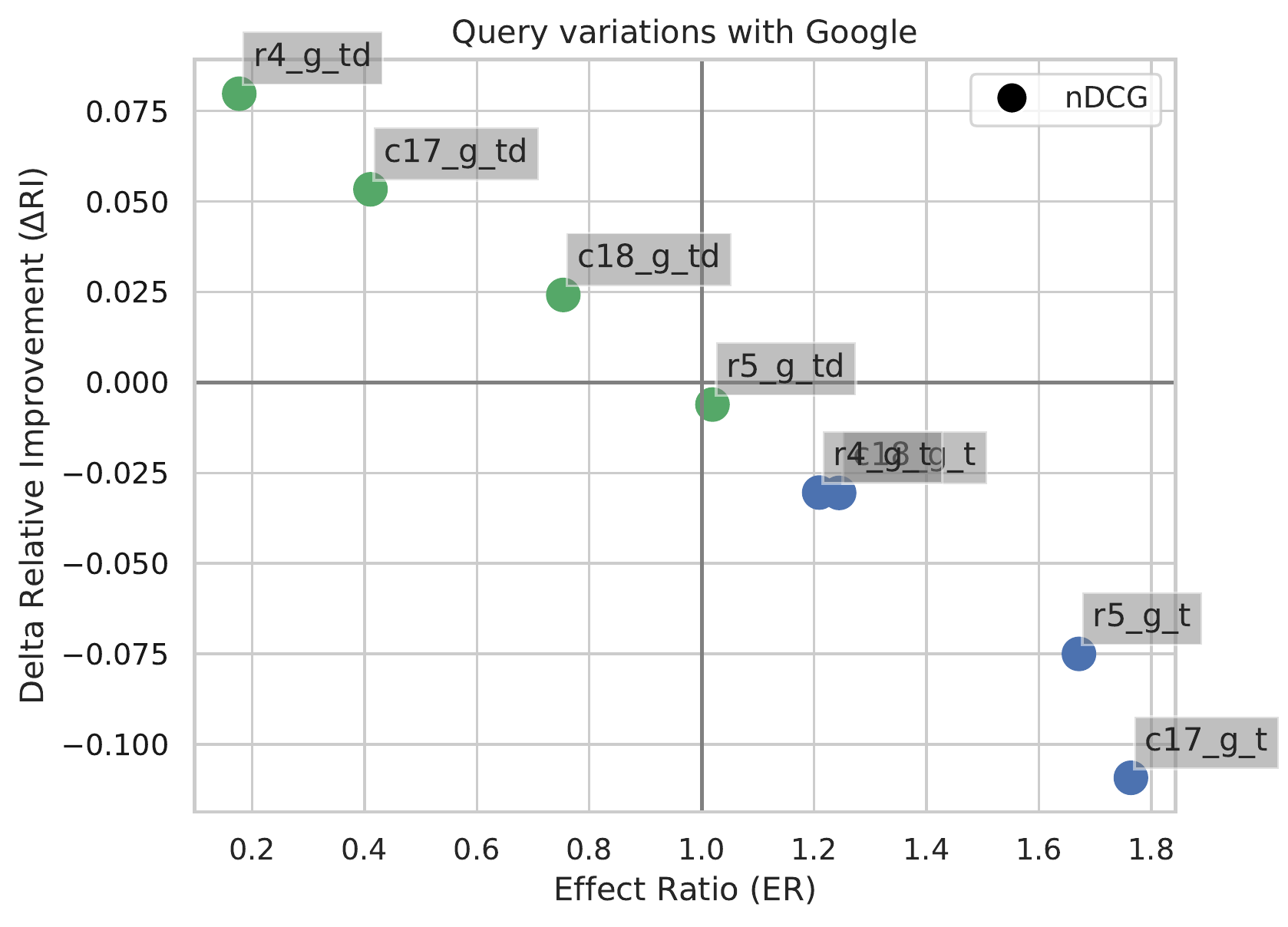}
    \includegraphics[width=5.55cm]{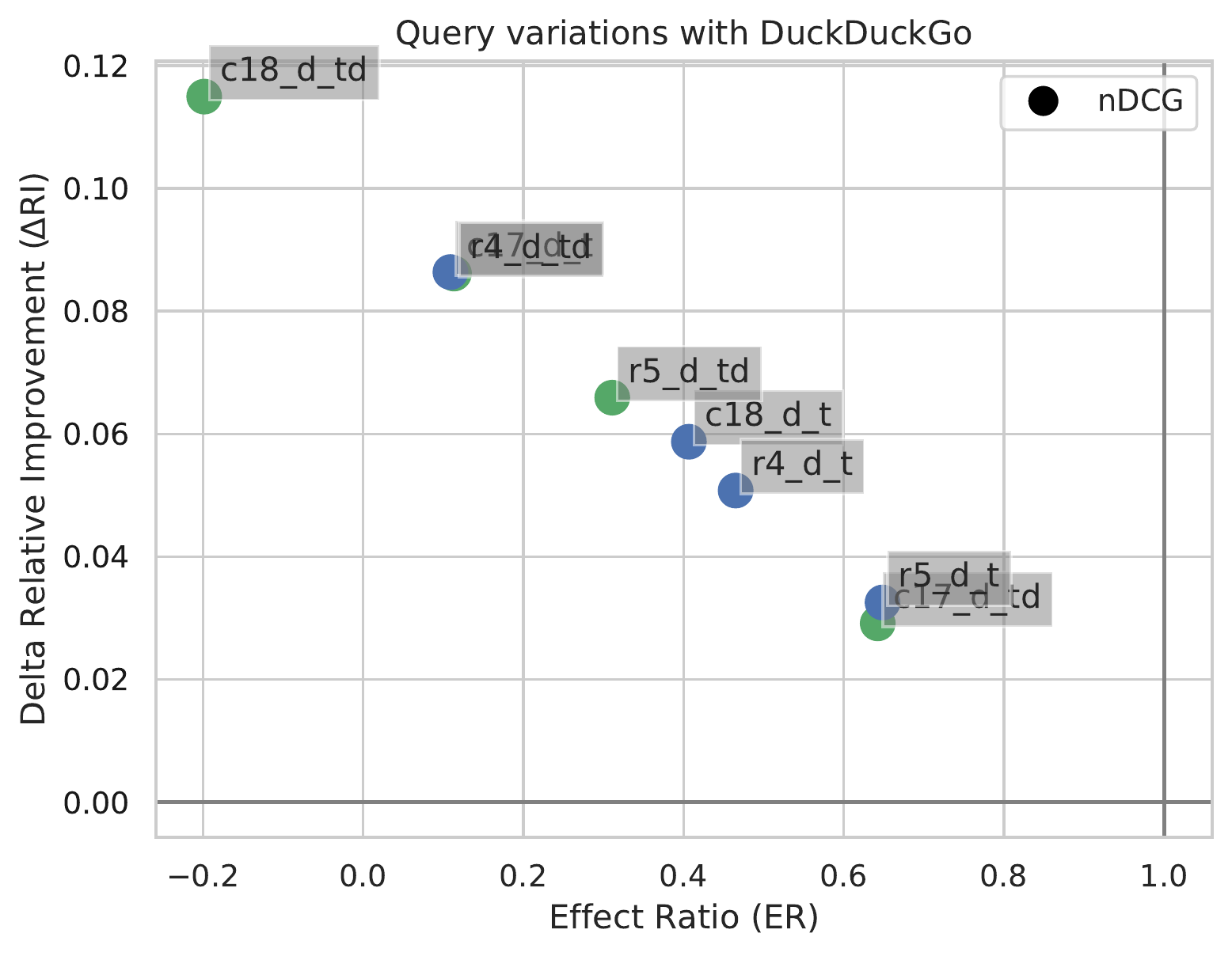}
\end{minipage}
\end{table}

\subsubsection{Further Discussions}

Referring back to our research questions \textbf{RQ1} and \textbf{RQ2}, we reflect on the influence of the targeted aspects with regard to the reproducibility measures provided by \texttt{repro\_eval} (KTU, RBO, RMSE, ER, and DRI). \textbf{RQ1} addresses the change of retrieval performance over time under the consideration of possibly different search engines and query formulations. Even though the experiments showed clear differences between the orderings of documents  and  the topic score distributions after two years, no substantial differences in average retrieval performance (ARP) are present. Even the more granular investigations of the temporal influence at intervals of two days (cf. Figure~\ref{fig:time_series}) showed that the performance is robust and is independent of individual SERP ranking lists.

With regard to the \textit{web search engine}, our reimplementations delivered higher baseline scores when training data is retrieved from snippets of DuckDuckGo-SERPs for each test collection. Especially for title-only queries, there are significant differences in comparison to the runs derived with Google results. Due to increased baseline scores, the overall effects were lower for DuckDuckGo than those runs based on Google results. At the level of overall effects, we can clearly distinct the results from two different web search engines, especially if the training data is retrieved from Google with title-only queries, where the overall effects are much higher (cf. Table~\ref{tab:overall_effects}). When we conducted our experiments, DuckDuckGo had longer snippet texts that may lead to more expressive training data. We leave it as future work to investigate the interactions and effects between the query and the snippet length (volume of the training data). 

How do the two different \textit{query formulations} affect the final run results?  In most cases, querying web search engines with titles only results in lower scores for the baseline runs than queries made from topic titles and descriptions. While it is a common finding, that retrieval performance can benefit from concatenation of titles and descriptions, e.g., as already shown by Walker et al.~\cite{DBLP:conf/trec/WalkerRBJJ97}, it is interesting to see that these effects ``carry over'' in this specific setup, where the queries rather affect the quality of training data and are not directly used to derive the runs from the test collection.

\textbf{RQ2} addresses the extent to which the original effects can be reproduced in different contexts with other newswire test collections. Comparing the ARP with different test collections does not show significant differences, which indicates that the procedure is robust in terms of ARP. It is impossible to compare some aspects with a different test collection, i.e., the ordering of documents and the topic score distributions cannot be directly compared. The ER and DRI measures are proxies that compare the effects between the baseline and advanced run. Using Google, shorter queries lead to stronger effects than in the original experiment. % between the baseline and the advanced run than in the original experiment.
On the contrary, the resulting effects based on longer queries with Google and DuckDuckGo (with both query types) stay below those of the original experiments. This is consistent for all test collections.

\section{Conclusion}
We analyzed the topic-specific data enrichment of a pseudo-relevance method that is based on web search results. Motivated by Grossman and Cormack's submissions to TREC Common Core 2018, we reimplemented the original workflow and analyzed different influencing factors related to the web search that affect the constitution of the data enrichment, i.e., the training data. Our experiments demonstrate the influence of the web search engine and the query formulation. Even though the composition of SERPs (which are the foundation of the training data) changes over time, the average retrieval performance is not affected, and the results are robust. This shows that SERP snippets and linked web page content can be reliably used as an external corpus for the investigated ranking method. Furthermore, we analyzed the experiments in different contexts with other newswire test collections. In our experiments, we did not consider other elements of SERPs that might contribute to more effective retrieval results. It is of future interest to consider more targeted ways to extract texts from web pages or SERPs that improve the quality of the training data and investigate the influence of the classifier by replacing it with more sophisticated deep learning approaches. Besides our open-source reimplementation, we also provide the scraped artifacts and system runs in a public Zenodo archive\footnote{\href{https://doi.org/10.5281/zenodo.4105885}{\url{https://doi.org/10.5281/zenodo.4105885}}}.

\paragraph{Acknowledgments} This paper is supported by the DFG (project no. 407518790).

% \bibliography{bibliography} 
\bibliography{short} 
\bibliographystyle{acm}
\end{document}